\documentclass[aps,prl,twocolumn,groupedaddress]{revtex4}

\usepackage{graphicx}
\usepackage{amsmath,amssymb}
\usepackage{color}
\usepackage{setspace}

\begin{document}


\title{Violation of a temporal Bell inequality for single spins in solid \\by over 50 standard deviations}


\author{G. Waldherr$^1$}
\email{g.waldherr@physik.uni-stuttgart.de}
\author{P. Neumann$^1$}
\author{S.~F. Huelga$^2$}
\author{F. Jelezko$^{1,3}$}
\author{J. Wrachtrup$^1$}
\affiliation{$^1$ 3. Physikalisches Institut and Research Center SCOPE, University of Stuttgart, Pfaffenwaldring 57, 70569 Stuttgart, Germany}
\affiliation{$^2$ Institut f\"{u}r Theoretische Physik, Universit\"{a}t Ulm, Ulm D-89073, Germany}
\affiliation{$^3$ Institut f\"{u}r Quantenoptik, Universit\"{a}t Ulm, Ulm D-89073, Germany}


\date{\today}

\begin{abstract}

Quantum non-locality has been experimentally investigated by testing different forms of Bell's inequality, yet a loophole-free realization has not been achieved up to now.
Much less explored are temporal Bell inequalities, which are not subject to the locality assumption, but impose a constrain on the system's time-correlations.
In this paper, we report on the experimental violation of a temporal Bell's inequality using a nitrogen vacancy defect (NV) in diamond and provide a novel quantitative test of quantum coherence.
We therefore present a new technique to initialize the electronic state of the NV with high fidelity, a necessary requirement for reliable quantum information processing and/or the implementation of 
protocols for quantum metrology.
\end{abstract}


\maketitle

Ever since the birth of quantum mechanics, its non-determinism and non-local structure have motivated increasing theoretical scrutiny. 
More recently, discussions have moved to the experimental realm but many aspects of the initial debate still remain to be fully understood. 
The superposition principle underpins quantum non-determinism and its application to systems of increasing complexity can lead to contradictions with our direct experience of the world, as neatly exemplified by the Schr\"odinger's Cat Gedankenexperiment \cite{Schroedinger1935}.  
Schr\"odinger envisaged a (quantum) cat trapped inside a box equipped with some device which may, randomly, kill the cat.
Quantum mechanics tells us that at any time, if unobserved, the cat is both dead and alive.
However, if the system is probed, for instance by opening the box, we will find the cat in one of the two possible states, either alive or dead.
According to the formalism of quantum mechanics, this result is non-deterministic, i.e. the outcome of our measurement, and therefore the (classical) state of the cat is not even defined before opening the box.
But how can we know that quantum mechanics is not simply incomplete, and that there are not other, ``hidden'' variables, on which the state of the cat depends on, thus rendering the measurement deterministic?
When formulated in the context of spatially separated subsystems, this type of argument leads to the celebrated Bell's inequality, which limits the strength of the statistical correlations between distant subsystems within a local realistic theory \cite{Bell1964, Clauser1969, Aspect1982, Greenberger1990}.

On the other hand, in contexts where locality is not relevant, the focus is on the premises of {\em realism} and the characterization of the type of temporal (vs spatial) correlations that would emerge within that realist description. This issue was first investigated by Leggett and Garg in 1985 \cite{Leggett1985} and led to the formulation of the so called {\em temporal} Bell inequalities (TBI).
The major difference to the original Bell inequalities is that instead of correlations between states of two spatially separated systems, one is now concerned with correlations of the state of a single system at different points in time.
In \cite{Leggett1985}, a measurable inequality is obtained from the assumptions of {\em macroscopic realism} (a macroscopic system is always in one of its macroscopically distinct states) and the possibility to perform {\em non-invasive measurements} (measurements do not influence the dynamics of the system).
It is clear that these assumptions, and therefore the resulting inequality, are typically violated by quantum mechanics. However, we would expect these assumptions to become valid at some point as the considered system becomes closer and closer to a truly macroscopic object.
Within this view, TBI provide a criterion to characterize the boundary between the quantum and classical domains and the possible identification of macroscopic quantum coherence.
Leggett and Garg suggested an experimental scenario for testing their inequality using superconducting interferometric devices (SQUIDs), whose macroscopically distinct states would correspond to counter-propagating currents.
Other proposals for experimental tests followed \cite{Tesche1990, Paz_PRL1993}, but the actual experimental implementation was hindered by the assumption of non-invasive measurability, which is violated by projective measurements.
There were nevertheless some suggestions to circumvent this by using a measurement strategy, which, together with the realism assumption, would result in non-invasive measurements, e.g. a ``negative result measurement'' as suggested by Leggett and Garg, or delayed-choice measurements \cite{Paz_PRL1993}.
A more recent approach is to use weak measurements, assuming that the dynamics of the system (including possible hidden variables) is thereby disturbed only very slightly.
Using this method, the first experimental violation of temporal Bell inequalities was recently demonstrated in superconducting qubits \cite{Palacios_NatPhys2010}.

Here we will follow a different approach and use the temporal Bell inequality presented in \cite{Huelga_PRA1995,Huelga_PRA1996} based upon the strategy of replacing the condition of non-invasive measurements by an additional assumption of stationarity of the correlations.
The advantage of this formulation is that it leads to easily testable inequalities (using projective von Neumann measurements), whose violation can be interpreted in a transparent way.
Specifically, we will test a temporal Bell inequality based upon the following assumptions:
\begin{description}
\item[] A1) ``Reality'': The state of any physical system is always well defined, i.e. the dichotomic variable $M_i(t)$, which tells us whether ($M_i(t) = 1$) or not ($M_i(t) = 0$) the system is in state $i$, is, at any time, $M_i(t) = {0, 1}$.
\item[] A2) ``Stationarity'': The conditional probability $Q_{ij}(t_{1}, t_{2})$ to find a system in state $j$ at time $t_{2}$, if it was in state $i$ at time $t_{1}$ only depends on the time difference $t_{2}-t_{1}$.
\end{description}
Effectively, the stationarity assumption implies restrictions on possible deterministic hidden variable theories, which cannot be ruled out by experimental violation of the inequality (cf. \cite{Huelga_PRA1996} for details).
However, the assumed stationarity of any two-time correlation function is in principle amenable to separate experimental testing and expected to hold not only for idealized closed quantum dynamics but also in open systems subject to purely Markovian noise at a rate $\gamma$, where idealized sinusoidal two-time correlations would be exponentially damped with a factor $\gamma (t_i-t_j)$ \cite{Breuer_Book}.
In a real experiment, a non-stationary dynamics could arise as a result of possible non-Markovian interactions.
In our case, the environmental back-action of the environment onto the system is negligibly small in the observed time scales and the considered parameter regimes, which endorses the validity of the stationarity assumption in this scenario and therefore enhances the physical character of the hidden variables theories put to test.

Following \cite{Huelga_PRA1996}, the reality and the stationarity assumptions yield the constrain
\begin{equation}
Q_{ii}(0,2t)-Q_{ii}^{2}(0,t) \geq 0 \label{eq:tbi}.
\end{equation}
This inequality bounds the strength of the temporal correlations that can arise in a classical framework and, like the original TBI, can be violated by quantum mechanical unitary dynamics, e.g. by a Rabi oscillation or Larmor precession.
When the system's evolution is not closed, violation of the classical bound can persist depending on the noise strength.
For the case of Markovian noise, above a certain value of the noise intensity and after certain time, the dynamical evolution will not longer violate the classical limit and the observed conditional probabilities could be accounted for in terms of statistical mixtures of orthogonal states.
On the other hand, when the classical limit is violated, the system has a degree of coherence leading to observable probabilities that cannot be simulated without resorting to superposition of orthogonal states, which are alien to a realist description of the dynamics.
Within this view point, an experimental test of the inequality (\ref{eq:tbi}) provides supplementary, quantitative information to traditional schemes to test for quantum coherence in terms of Ramsey fringes, given that it sets a maximal loss of visibility of the fringes pattern in a given time interval in order to violate the classical bound and therefore ensure the impossibility of an alternative hidden variables description.

\begin{figure}[tbp]
	\centering
		\includegraphics[width=0.45\textwidth]{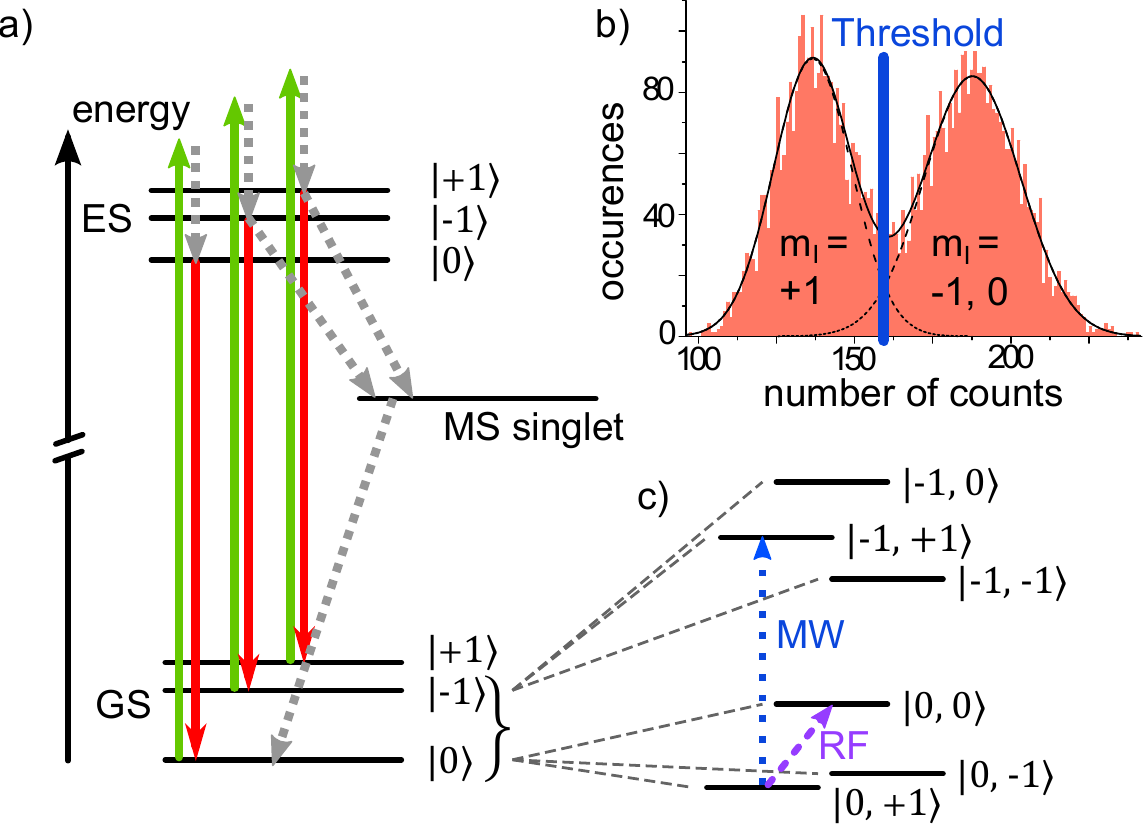}
	\caption{a) Energy level scheme of the NV$^-$ defect in a small magnetic field. Optical transitions occur between ground state (GS) and excited state (ES) (vertical arrows), dotted lines (gray) indicate radiationless decays. Line thickness corresponds to transition rates. b) Histogram of many subsequent QND measurements of the nitrogen nuclear spin. Low fluorescence level indicates that the MW $\pi$ pulse was successful, i.e. that the nuclear spin state is $m_I = +1$. c) Magnification of the $m_S = 0, -1$ levels including hyperfine splitting due to the $^{14}N$ nuclear spin (states are denoted as $|m_S, m_I\rangle$). The dotted (blue) arrow illustrates the nuclear spin selective MW $\pi$ pulse, the dashed (purple) arrow the nuclear spin transition driven by RF pulses. }
	\label{fig:NVLevels}
\end{figure}

We now describe the experimental set up and the obtained results.
In order to measure the conditional probabilities in the TBI (\ref{eq:tbi}), we need projective quantum non demolition (QND) measurements.
We will use the nitrogen nuclear spin associated with the nitrogen-vacancy defect (NV) in diamond to experimentally test this inequality, using repetitive readout \cite{Jiang_Science2009} and high magnetic fields that allows ``single-shot'' measurements \cite{Neumann_Science2010}.
Another approach to QND measurements of the NV is dispersive coupling to light \cite{Buckley_Science2010}.
The basic level scheme of the negatively charged NV (NV$^{-}$) is depicted in fig. \ref{fig:NVLevels} a).
Green light brings NV$^{-}$ into its  excited state (ES) while conserving the spin polarization, from where it has two different decay paths depending on the electronic spin state.
For the $m_S = 0$ state there will be predominantly radiative decay into the $m_S = 0$ ground state (GS).
On the other hand, if the spin state is $m_S = \pm 1$, an inter-system crossing to the metastable (MS) singlet state is more likely to occur, from where it decays into the $m_S = 0$ ground state.
Due to the (mostly) non-radiative nature of this decay, the fluorescence of the $m_S = \pm 1$ state is lower than from the $m_S = 0$ state.
This dynamics allows optical spin state detection, and also polarizes the NV$^{-}$ into the $m_S = 0$ ground state.

Whereas the electron spin state is destroyed during this process, the nuclear spin state can be made very robust by applying a strong magnetic field \cite{Neumann_Science2010} ($B = 0.6$ T in our case).
The nuclear spin state can be mapped onto the electron spin with a CNOT-gate (realized by a nuclear spin selective $\pi$ pulse onto the electron spin, cf. blue arrow in fig. \ref{fig:NVLevels} a)), which is then optically measured.
This quantum non-demolition (QND) measurement is repeated many times ($\approx$2000) to gather enough information to directly determine the nuclear spin state, by summing up the fluorescence photons.
Fig. \ref{fig:NVLevels} b) shows a histogram of many such readouts, showing two separate Poissonian distributions, corresponding to nuclear spin state $m_I = +1$ (left distribution) and $m_I = -1, 0$ (right distribution).
By introducing a threshold between the two distribution, the number of collected photons can be directly converted into nuclear spin state.
The overlap of the distributions introduces false results, decreasing the fidelity $F$, which is $F^2\approx0.91$ for initialization + readout in our case.

All these measurements are carried out on NV$^{-}$.
As has recently been shown \cite{Waldherr2011}, under typical measurement conditions the NV resides ~30\% of time in its neutral state NV$^{0}$.
This is due to a two-photon-ionization process during illumination, at which the first photon brings the NV$^{-}$ into its excited state, from where it is ionized to NV$^{0}$ by the second photon.
Illumination with green light also restores NV$^{-}$.
Although QND readout of the nuclear spin is possible under this condition, radio-frequency (RF) pulses aimed for nuclear spin transitions in the NV$^{-}$ $m_s=0$ ground state will have no effect on NV$^0$ (30\% of times), because of different hyperfine splitting.
Additionally, coherent driving in NV$^0$ is hardly possible due to fast dephasing.This hinders the observation of high contrast Rabi oscillation, necessary to show violation of the TBI (\ref{eq:tbi}).

\begin{figure}[tbp]
	\centering
		\includegraphics[width=0.45\textwidth]{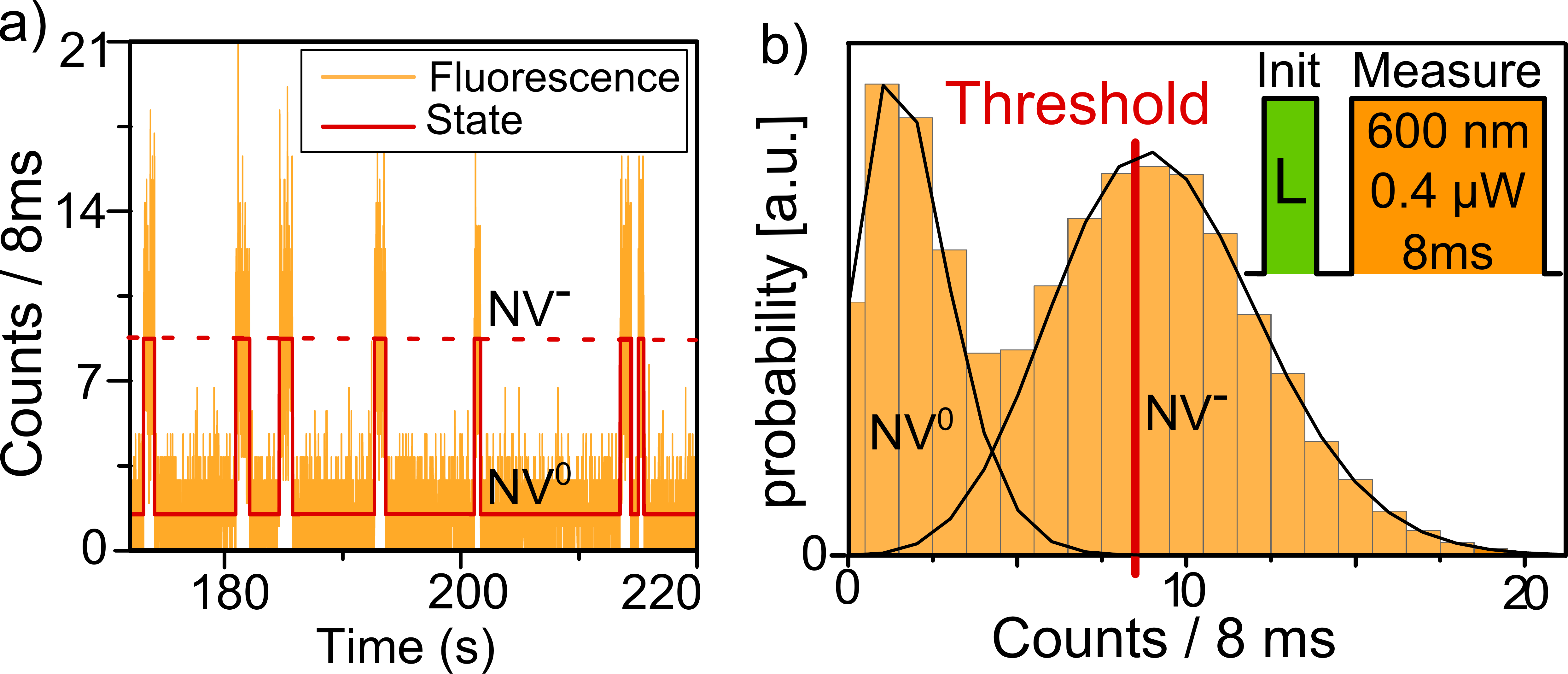}
	\caption{a) Timetrace of the fluorescence of the NV under cw illumination with 600nm, 0.4$\mu$W laser light (orange line).
	The red line shows the most likely fluorescence level.
	The lifetime of NV$^{-}$ during orange illumination is around 600 ms.
	b) Histogram of measurement results showing the distribution of counted photons during orange illumination for the inset sequence: Green pulses ``reset'' the NV and 8 ms orange pulses measure the charge state. 
	}
	\label{fig:NVSwitch}
\end{figure}

In the following, we present a new technique to non-destructively detect and thereby initialize the charge state of the NV.
Since the zero phonon line (ZPL) of NV$^{0}$ is 575nm, illumination of light with wavelength $\lambda \gg 575$nm will not induce any fluorescence in this state.
On the other hand, fluorescence of the NV$^{-}$ can be observed for illumination wavelengths of up to 637nm, which is the ZPL of NV$^{-}$.
The ionization rate of NV$^{-}$ decreases quadratically with the illumination power (two photon process) \cite{Waldherr2011}, whereas the fluorescence only decreases linearly.
Using low power excitation light, it is therefore possible to measure fluorescence from NV$^{-}$ before it becomes ionized.
This allows to initialize the NV in its negative charge state, after the initialization of the nuclear spin.
Fig. \ref{fig:NVSwitch} a) shows a timetrace of the fluorescence of the NV under illumination with low power ($~0.4\mu$W, cw) orange light ($\lambda = 600$nm).
Fig. \ref{fig:NVSwitch} b) shows a histogram of charge state measurements after a green laser pulse (typically used to measure the spin state of NV$^-$).
The two different fluorescence state corresponding to NV$^{0}$ (low fluorescence) and NV$^{-}$ (high fluorescence) are clearly distinguishable by setting a threshold in the middle between the two peaks.
Since we only want to assure that the NV is in its negative charge state, this threshold can be shifted towards higher count rates to increase the fidelity.
Only if we find the NV to be in its negative charge state, we use the result of the subsequent experiment.

\begin{figure}[tbp]
	\centering
		\includegraphics[width=0.45\textwidth]{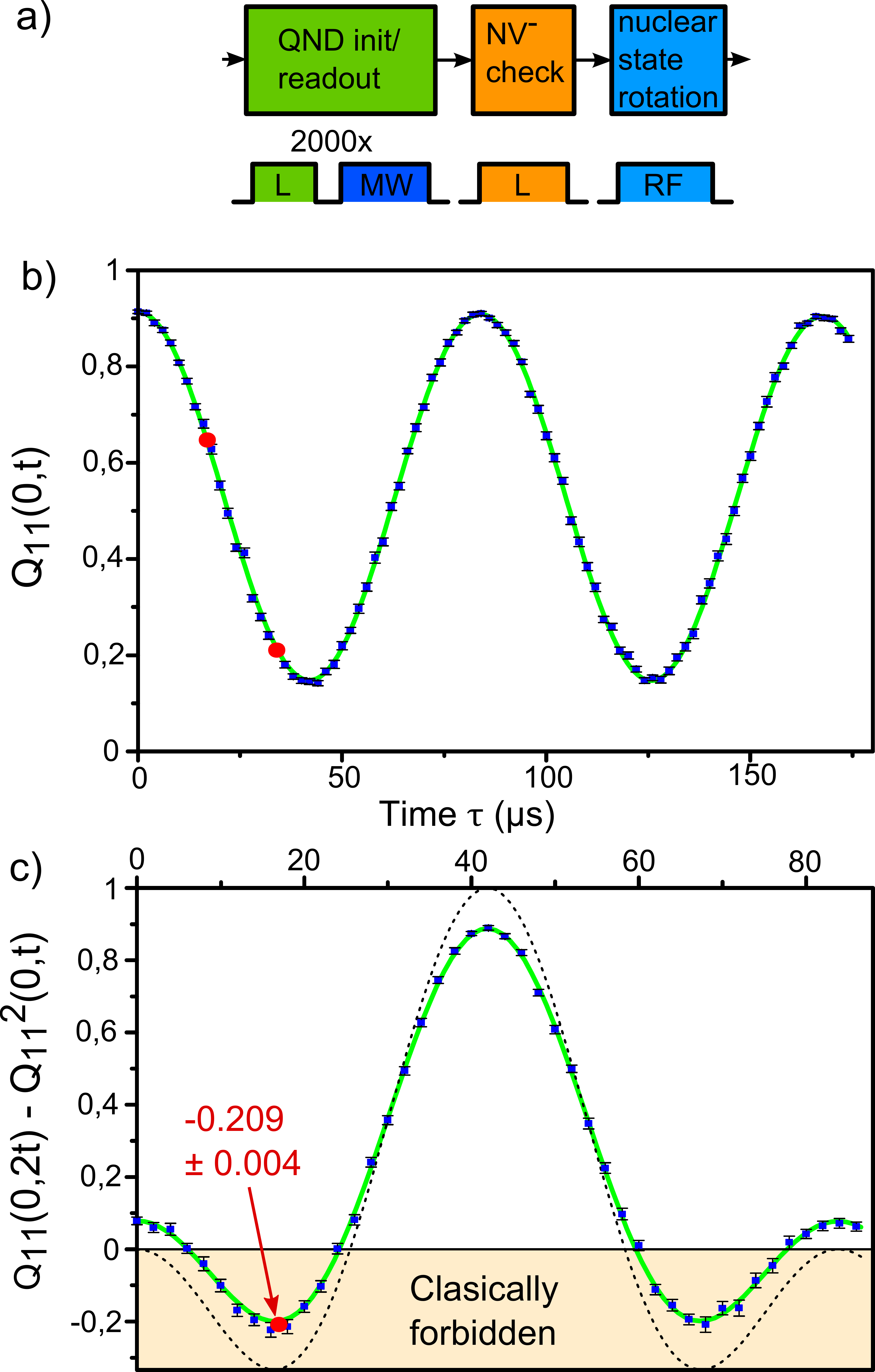}
	\caption{a) Measurement sequence (cf. text). b) Rabi oscillation of the nuclear spin, introduced by RF-pulses of varying length $\tau$ (time-axis), the blue line shows a cos-fit. Red dots indicate points measured with higher accuracy. c) Result of the temporal Bell inequality (\ref{eq:tbi}). The green line is obtained by calculating the inequality for the fit curve of b), the dotted black line is for comparison and shows the "`ideal"' case (error-free measurement of Rabi oscillation). The red dot corresponds to the red dots in b).}
	\label{fig:rabi}
\end{figure}

A nuclear Rabi oscillation obtained with this method is shown in fig. \ref{fig:rabi} b).
The curve is slightly shifted upwards, i.e. in some times the NV is not in the $m_S=0$ NV$^-$ ground state (cf. \cite{Waldherr2011}).
This can be attributed to the limited fidelity of the charge state measurement, as well as to imperfect polarization of the electron spin, partly due to $T_1$ decay during the charge state measurement.
The error bars show empirical errors calculated from many sub ensembles of the measurements for the respective point.

The measurement procedure for the Rabi oscillation using single-shot readout is the following (cf. fig. \ref{fig:rabi} a)):
i) Initializing the nuclear spin by measuring its state,
ii) Measuring the charge state of the NV,
iii) The application of radio-frequency (RF) pulses for time $\tau$ to rotate the nuclear spin state on the Bloch sphere,
iv) Measuring the final state of the nuclear spin. This measurement will always yield either 1 or 0, no events are missed as is the case with single photon detection \cite{Rowe_Nature2001}.
Averaging over many such measurements directly yields the conditional probabilities $Q_{11}(0,t), Q_{11}(0,2t)$, readily derived from the Rabi oscillations. 
The result of the calculation of the temporal Bell inequality \ref{eq:tbi} is shown in fig. \ref{fig:rabi} c), demonstrating a clear violation of the inequality.
The red dot in fig. \ref{fig:rabi} c), corresponding to the two red dots in the Rabi oscillation (fig. \ref{fig:rabi} b)), gives the largest violation of eq. \ref{eq:tbi} and was measured with very high accuracy.
Its value is $-0.209 \pm 0.0039$, which is over 50 standard deviations below 0.

This result demonstrates for the first time, that the dynamics of a single nuclear spin cannot be described by a realist theory supplemented with hidden variables.
In future experiments, by entangling two nuclear spin suitable for single-shot readout, also the contextuality of quantum mechanics could be experimentally tested \cite{Kirchmair_Nature2009}.
For future application such as quantum information processing and Heisenberg limited phase estimation \cite{Higgins_Nature2007, Higgins_NewJPhys2009}, refining experiments ins terms of stabilizing the NV in its negative charge state will be of pivotal importance.
Otherwise, the NV resides ~30\% of time in its neutral charge state, which does not offer the unique, favorable properties of NV$^-$ \cite{Felton2008, Waldherr2011}.

\bigskip
We are grateful to Martin Plenio and Emilio Santos for their comments. 
Support by the European Commission (Integrated projects Q-Essence, SOLID and the STREPs, 
PICC, CORNER, SQUTEC, and DIAMANT) and the Deutsche Forschungsgeminschaft (SFB/TR 21) is gratefully acknowledged.

\bibliography{jebref_database}

\end{document}